\documentclass[showpacs,twocolumn,superscriptaddress]{revtex4-2}
\usepackage{graphicx}
\usepackage{physics}
\usepackage{wrapfig}
\usepackage{changepage}
\usepackage{amsmath,amssymb,amsthm,mathrsfs,amsfonts,dsfont,mathtools}
\usepackage[caption=false]{subfig}
\usepackage{epsfig}
\usepackage{color,xcolor}
\usepackage{adjustbox}
\usepackage{tabularx}
\usepackage{array}
\usepackage{multirow}
\usepackage{tabu}
\usepackage{bm}
\usepackage{enumerate}
\usepackage{hyperref}
\usepackage{newtxtext,newtxmath}
\usepackage{makecell}%To keep spacing of text in tables
\usepackage{url}
\usepackage{booktabs}
\usepackage{nicefrac}%\biboptions{sort&compress}
\usepackage{makeidx}  % allows for indexgeneration
\usepackage{algorithm}
\usepackage{algpseudocode}
\usepackage{cleveref}
\usepackage{mwe}

\usepackage{comment}

\begin{document}

\title{Scalable quantum measurement error mitigation via conditional independence and transfer learning}

\author{ChangWon Lee}
%\email{changwonlee@yonsei.ac.kr}
\affiliation{Department of Statistics and Data Science, Yonsei University, Seoul 03722, Republic of Korea}
\author{Daniel K. Park}
\email{dkd.park@yonsei.ac.kr}
\affiliation{Department of Statistics and Data Science, Yonsei University, Seoul 03722, Republic of Korea}
\affiliation{Department of Applied Statistics, Yonsei University, Seoul 03722, Republic of Korea}

\begin{abstract}
Mitigating measurement errors in quantum systems without relying on quantum error correction is of critical importance for the practical development of quantum technology. Deep learning-based quantum measurement error mitigation has exhibited advantages over the linear inversion method due to its capability to correct non-linear noise. However, scalability remains a challenge for both methods. In this study, we propose a scalable quantum measurement error mitigation method that leverages the conditional independence of distant qubits and incorporates transfer learning techniques. By leveraging the conditional independence assumption, we achieve an exponential reduction in the size of neural networks used for error mitigation. This enhancement also offers the benefit of reducing the number of training data needed for the machine learning model to successfully converge. Additionally, incorporating transfer learning provides a constant speedup. We validate the effectiveness of our approach through experiments conducted on IBM quantum devices with 7 and 13 qubits, demonstrating excellent error mitigation performance and highlighting the efficiency of our method.
\end{abstract}
% \keywords{Quantum error mitigation, deep learning, conditional independence, transfer learning}
\maketitle

%########### body
\section{Introduction}
\label{sec:section1}
%Quantum computing has emerged as a promising technology with the potential to revolutionize various fields, including cryptography, optimization and financial modelling. 
Quantum computing offers computational advantages over classical algorithms across various problems, such as factoring integers, simulating quantum systems, solving linear systems of equations, machine learning, and simulating stochastic processes~\cite{shor1999polynomial, feynman2018simulating,peruzzo_variational_2014,McClean_2016,harrow2009quantum,QSVM,QPCA,montanaro2015quantum,PhysRevLett.125.260501,blank_quantum-enhanced_2021}. However, the susceptibility of quantum computing to noise and imperfections poses a significant challenge, limiting its ability to surpass classical capabilities in solving real-world problems. While the theory of quantum error correction (QEC) and fault-tolerance holds the promise of scalable quantum computation~\cite{aharonov1997fault,fowler2012surface}, building a fault-tolerant quantum computer remains a long-term endeavor. In the ongoing efforts to build full-fledged fault-tolerant quantum computers, there is a desire for techniques that improve the utility of quantum hardware in the presence of noise without relying solely on QEC.

Quantum error mitigation (QEM) refers to a set of techniques aimed at reducing the impact of errors on the outcomes of quantum computations~\cite{temme2017error,endo2018practical,9226505,kurita2022synergetic}. Unlike QEC, which completely removes errors, QEM focuses on minimizing their effects on the final result of an algorithm. By relaxing the requirement for full recovery of the desired state, QEM techniques can be implemented without the need for additional physical qubits. This makes QEM particularly well-suited for near-term quantum computing, where the size of quantum circuits that can be reliably executed is limited. In fact, QEM plays a crucial role in the Noisy Intermediate-Scale Quantum (NISQ) era~\cite{preskill2018quantum}, as it maximizes the utilization of limited quantum resources and expands the capacity of quantum systems for solving real-world problems~\cite{li2017efficient,Qutility}.  In this respect, developing the most efficient and scalable QEM techniques is an important task.
% One advantage of QEM is that it does not require any additional physical qubits. Instead, QEM applies post-processing techniques to the data obtained from the quantum computer to reduce errors, which can help improve the accuracy and reliability of NISQ computations by leveraging limited quantum resources. There are many different QEM methods. For examples, Zero-noise extrapolation (ZNE), Probabilistic error cancellation (PEC), Measurement error mitigation(MEM)~\cite{endo2018practical, temme2017error, chen2019detector, bravyi2021mitigating, maciejewski2021modeling}. This paper focuses on quantum error mitigation techniques designed for addressing measurement errors.

Measurement is an essential operation in quantum computing, but is prone to errors. In certain quantum devices, measurement errors can severely damage the overall computation. For instance, IBM quantum devices available on the cloud typically exhibit measurement error rates on the order of $1\%$, with some cases reaching as high as $40\%$. Various methods have been proposed to mitigate measurement errors~\cite{chen2019detector,Maciejewski2020mitigationofreadout,9142431,PhysRevApplied.17.014024,kim2022quantum}, all of which are based on fully characterizing the underlying noise model using techniques such as tomography and machine learning. However, the computational costs associated with these methods scale exponentially with the number of qubits, imposing limitations on both scalability and practicality.

In this paper, we present a scalable deep learning-based method for quantum measurement error mitigation (QMEM). Our method leverages the concepts of conditional independence and transfer learning~\cite{pan2010survey} to significantly improve the efficiency compared to previous methods. Conditional independence assumes that the impact of measurement cross-talk between distant qubits is negligible. This assumption is especially relevant for quantum devices with limited connectivity among physical qubits, such as those constrained by nearest-neighbor couplings~\cite{doi:10.1073/pnas.1618020114} or employing distributed modular architectures~\cite{PhysRevA.89.022317,PhysRevX.4.041041,10.1063/5.0082975,9951310,khait2023variational}. By incorporating this assumption, we are able to exponentially reduce the size of neural networks used for QMEM. Transfer learning assumes the existence of an error component that is shared across all qubits. This assumption facilitates a constant factor reduction in training time by effectively leveraging pre-trained models. 
% Proof-of-principle experiments conducted on IBM quantum devices with 7 qubits and 13 qubits show that the underlying assumptions hold, and our QMEM method effectively reduces the measurement error.
To validate our approach, we conducted proof-of-principle experiments on IBM quantum devices with 7 and 13 qubits. The results demonstrate that the underlying assumptions hold and affirm the effectiveness of our QMEM method in reducing measurement errors.
% A deep neural network (DNN) was trained on data from a quantum circuits whose measurement results were known. The input layer to the DNN for training is the observed probability distribution of all computational base measurements. The DNN is trained using a loss function that quantifies the error compared to an ideal probability distribution. The DNN learns the error from two distributions and reduces the error for the new data. Furthermore, we use DNN to learn the results measured on the large quantum circuit using conditional independence and transition learning.

The remainder of the paper is organized as follows. We begin by setting up the problems and reviewing the two common approaches of QMEM in Section~\ref{sec:section2}. Section~\ref{sec:section3} presents the theoretical framework of our work, describing how the concept of conditional independence and transfer learning techniques are incorporated into the proposed QMEM method. In Section~\ref{sec:section4}, we provide detailed instructions on how to implement the proposed QMEM methods and describe experiments conducted through the IBM quantum cloud service. This section also includes a comprehensive performance comparison between the proposed QMEM methods and existing methods. Conclusions are drawn in Section~\ref{sec:section5}, along with discussions on directions for future works and open problems.

\section{Background}
\label{sec:section2}
% QMEM
%In this section, we discuss the relevant background on QMEM methods and how to use transfer learning and conditional independence for scalability. 
Many experimental setups for both the quantum circuit model and quantum annealing use projective measurement in the computational basis to perform readout of a quantum state. Moreover, positive operator-valued measurements can be realized through the projective measurement with ancillary qubits~\cite{PhysRevA.100.062317,lee2023variational_a}. Therefore, our primary focus is the development of error mitigation techniques to enhance the projective measurement in the computational basis. An ideal measurement on $n$ qubits results in the probability distribution, which can be represented as a vector $\boldsymbol{p} = \lbrack p_1, p_2, ..., p_{2^n}\rbrack$.
% The ideal probability distribution of the measurement outcome is represented as a vector, $P_{ideal} = \{p(1), p(2), ..., p(n)\}$. 
However, the observed probability distribution in experiments deviate from $\boldsymbol{p}$ due to measurement errors. We denote the observed probability for each bitstring as $\hat{p}_i$ and the error map as $\mathcal{N}$ such that $\hat{\boldsymbol{p}} = \mathcal{N}(\boldsymbol{p})$. The goal of QMEM is to minimize the loss function, $D(\boldsymbol{p}, \hat{\boldsymbol{p}})$ where $D$ is a distance measure that quantifies the discrepancy between the true and observed probability distributions. 
%Mention this somewhere: Implementing QMEM on the probability distribution is relevant to the cloud-based quantum computing environment. Additional QMEM can be performed at the hardware level by directly improving the readout signals.

The linear inversion method (LI-QMEM) assumes a noise model $\mathcal{N}(\boldsymbol{p})=\boldsymbol{\Lambda} \boldsymbol{p}$ and aims to reconstruct the noise matrix $\boldsymbol{\Lambda}$ through tomography. It produces an error-mitigated probability vector $\tilde{\boldsymbol{p}}=\boldsymbol{\Lambda}^{-1}\hat{\boldsymbol{p}}$~\cite{chen2019detector,Maciejewski2020mitigationofreadout,9142431}. 
In contrast, QMEM can be performed by training a deep neural network $\mathcal{F}$ to approximate the inverse noise function $\mathcal{N}^{-1}$~\cite{PhysRevApplied.17.014024,kim2022quantum}. The trained neural network produces an error-mitigated probability vector $\tilde{\boldsymbol{p}}=\mathcal{F}(\hat{\boldsymbol{p}})\approx \mathcal{N}^{-1}(\hat{\boldsymbol{p}})=\boldsymbol{p}$. This approach, referred to as NN-QMEM, is capable of correcting non-linear errors, which is not possible with LI-QMEM~\cite{kim2022quantum}. However, both LI-QMEM and NN-QMEM suffer from scalability limitations as the memory and computation time grow exponentially with the number of qubits. Recent estimations suggest that the current classical computational resources can only handle NN-QMEM for quantum systems of up to 16 qubits~\cite{kim2022quantum}. This work focuses on overcoming the scalability limitation of NN-QMEM, since it can effectively correct non-linear errors.
% LI, NN - QMEM
% In this work, we will conduct experiments using two methods. The first method is called Linear Inversion based QMEM (LI-QMEM). We can assume that the transformation between ideal and observed probability vector can be written using a response matrix $\Lambda$. $\hat{P}_{observed} = \Lambda P_{ideal}.$ After obtaining the matrix $\Lambda$, the ideal probability vector can be calculated $\Lambda^{-1} \hat{P}_{observed}$.\\ The second method is called Neural Network based QMEM (NN-QMEM). This method uses classical deep learning techniques to approximate $\mathcal{N}^{-1}$. NN-QMEM outperforms LI-QMEM by detecting non-linear effects with higher accuracy. 
% Since the $n$-qubit circuit, the response matrix size is $2^n \times 2^n$, and the number of input and output nodes in the neural network is $2^n$. 

\section{Theoretical Framework}
\label{sec:section3}
\subsection{Conditional Independence}
\label{sec:sec3_sub1}
% CI
Conditional independence (CI) is a fundamental concept in probability theory. It plays an important role in probabilistic models, simplifying the structure of the model and enabling an efficient analysis of the relationships between variables~\cite{dawid1979conditional, koller2009probabilistic}. It is valuable when modeling a large set of variables, where directly representing the joint distribution becomes challenging or impractical. By utilizing conditional independence relationships between variables, the joint distribution can be decomposed into smaller, more manageable components. This decomposition allows for a more tractable representation and analysis of complex probabilistic models. To illustrate this, consider the example of two random variables, $X$ and $Y$. We define $X$ and $Y$ as independent if and only if $p(X, Y) = p(X)p(Y)$. Independence between these variables leads to a partitioning of the probability distribution into two parts. For instance, if $X$ and $Y$ each take $2^{10}$ values, the full joint distribution $p(X, Y)$ would involve $2^{20}$ probabilities. Nevertheless, assuming independence between $X$ and $Y$ enables the decomposition of the joint distribution into the product of the individual distributions $p(X)$ and $p(Y)$. This decomposition significantly reduces the number of required probabilities to just $2^{10} + 2^{10} = 2^{11}$. Moreover, the definition of conditional independence is as follows. Let $X$, $Y$, and $Z$ be random variables. We say that $X$ and $Y$ are conditionally independent given $Z$ if the joint probability of $X$ and $Y$ given $Z$ can be expressed as $p(X, Y | Z) = p(X | Z)p(Y | Z)$. This indicates that the dependence between $X$ and $Y$ can be accounted for solely through their relationship with $Z$.

%In NN-QMEM, the size of the neural network can be exponentially reduced by leveraging the concept of conditional independence. The definition of conditional independence is as follows. Consider random variables $X$, $Y$ and $Z$. We say that $X$ and $Y$ are conditionally independent given $Z$ if the joint probability of $X$ and $Y$ given $Z$ can be expressed as $p(X, Y | Z) = p(X | Z)p(Y | Z)$.
%, and write $X \indep Y | Z$. 
%Thus, given Z, the joint distribution of X and Y is factorized into the product of the marginal distribution of X and the marginal distribution of Y. 
%Figure 1 is an example of using conditional independence. 
\begin{figure}[t]
    \centering
    \subfloat[]{
        \includegraphics[width=0.12\textwidth]{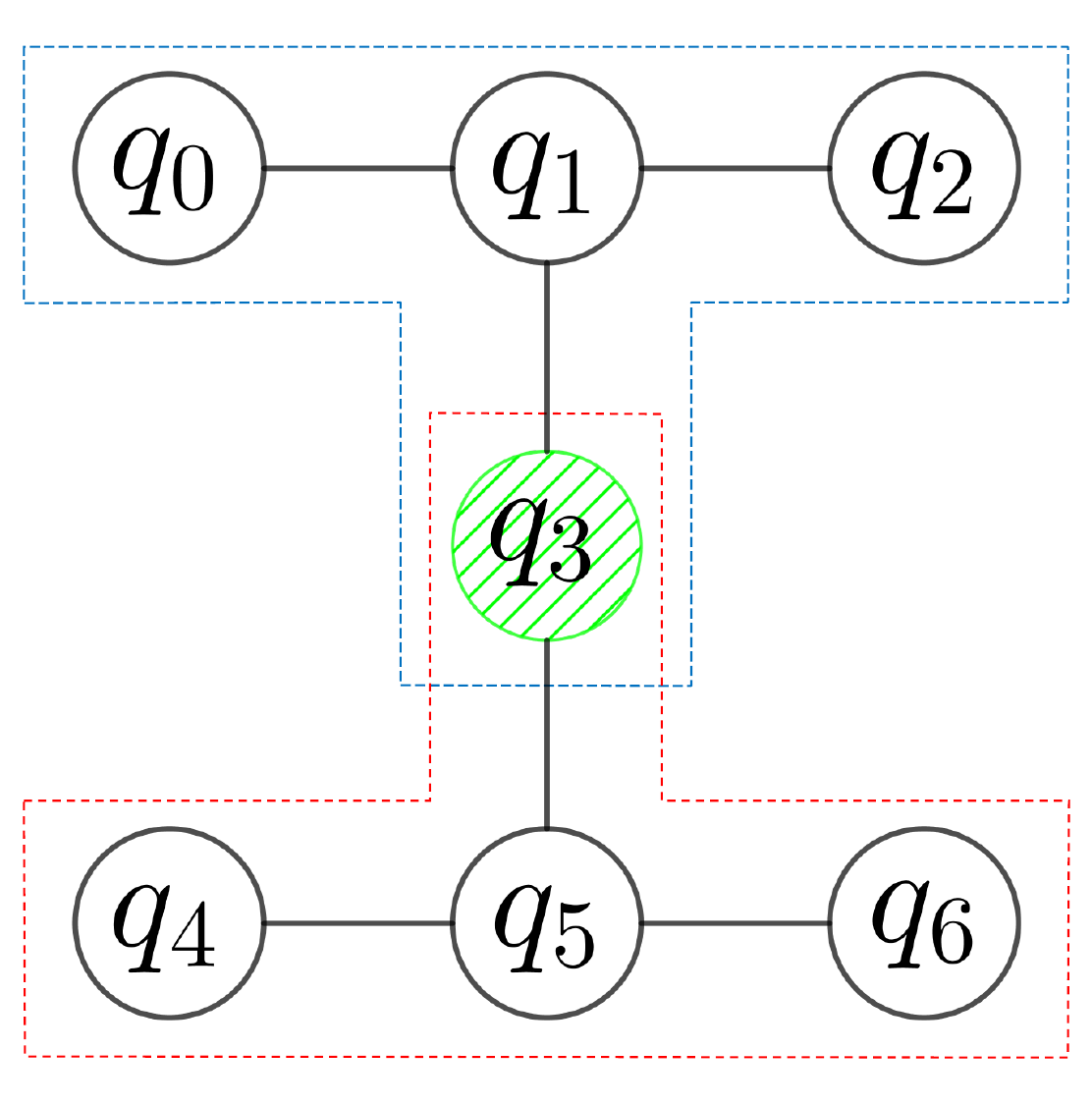}
        \label{fig:fig1_1}
    }
    \subfloat[]{
        \includegraphics[width=0.35\textwidth]{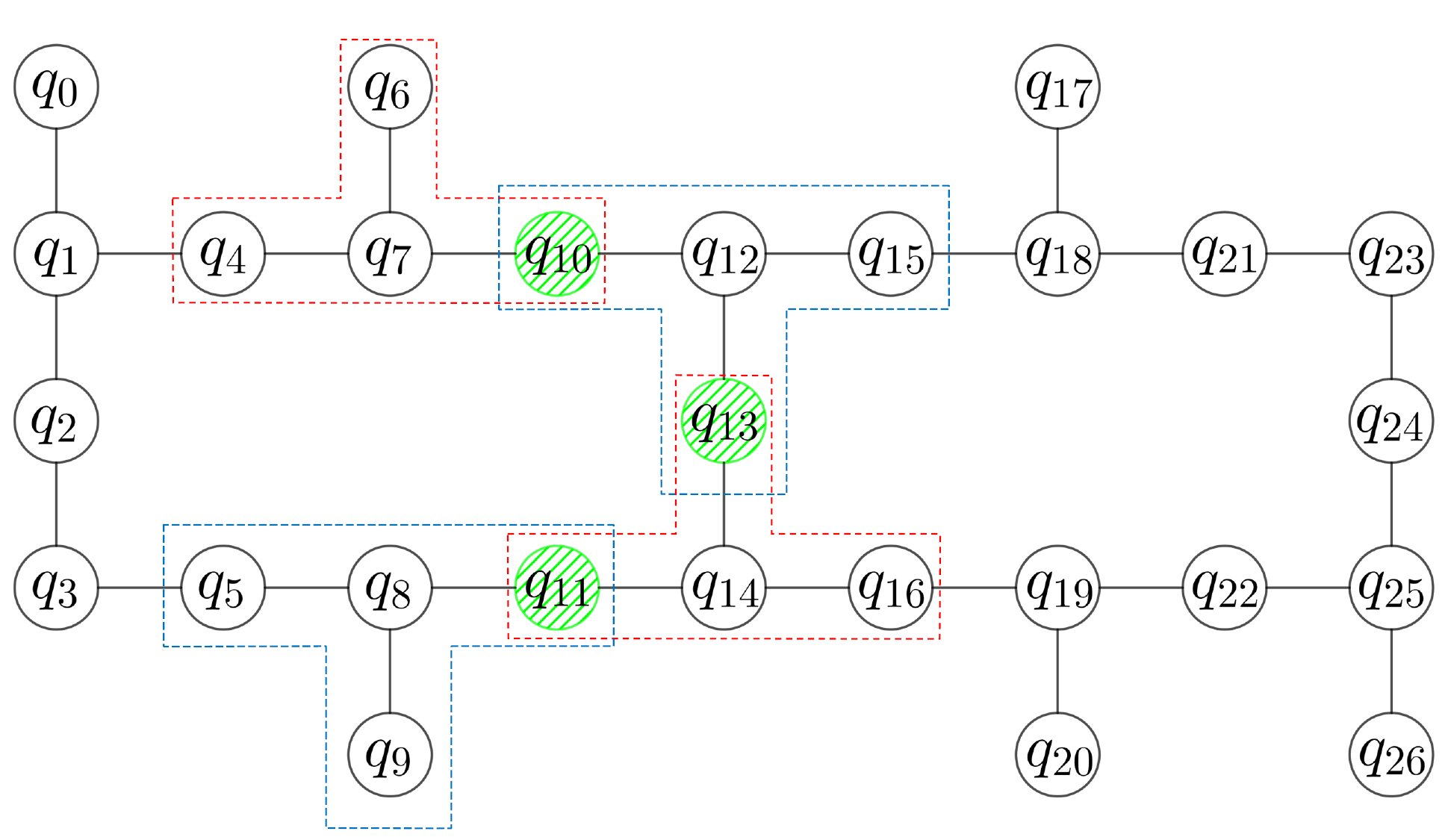}
        \label{fig:fig1_2}
    }
    \caption{\label{fig:fig1} Layouts of the IBM quantum devices used in this work: (a) 7-qubit device and (b) 27-qubit device. Each circle labeled with $q_i$ represents the $i^{\mathrm{th}}$ physical qubit on the devices, while the connecting lines represent the qubit connectivity. In (b), we selected 13 qubits (from $q_4$ to $q_{16}$) for the experiment. The shaded yellow qubits indicate conditional qubits, and the qubits grouped within T-shaped boundaries represent subsystems with their respective conditional qubits (i.e., the leaf node and its parent in Fig.~\ref{fig:fig2}). }
\end{figure}
To understand how the concept of CI can be applied to QMEM, let us consider a 7 qubit system depicted in Fig.~\ref{fig:fig1_1},  where $q_i$ denotes an $i^{\mathrm{th}}$ qubit. The circles and the wires in the figure represent physical qubits and their connectivity, respectively. The set of qubits $A=\lbrace q_{0},q_{1},q_{2}\rbrace$ and $B=\lbrace q_{4},q_{5},q_{6}\rbrace$ are connected only through $C=\lbrace q_{3}\rbrace$. Then assuming conditional independence of subsystems $A$ and $B$ given $C$, the joint probability distribution can be written as
\begin{align*}
    p(A, B, C)      = p(A, B | C) p(C)
                    = p(A | C) p(B | C) p(C).
\end{align*}
In the naive NN-QMEM, which aims to directly correct the full joint probability distribution $p(A,B,C)$, the number of input nodes of the neural network must grow exponentially with the number of qubits. Typically, the total number of nodes grows linearly with the number of input nodes, and the number of parameters grows quadratically. On the other hand, under the conditional independence assumptions, one needs three machine learning models that correct for $p(A | C)$, $p(B | C)$, and $p(C)$ independently. In this example, the number of input nodes for $p(A | C)$ and $p(B | C)$ is $2^3$, and is $2^{1}$ for $p(C)$. Since $C$ can take on two values (0 or 1), each conditional probability distribution requires two distinct neural networks. Consequently, the total number of parameters to be trained is proportional to $2((2^3)^2+(2^3)^2)+(2^1)^2=260$. In contrast, the full model requires it to be proportional to $(2^{7})^2 = 16384$. The reduced parameter count in neural networks also implies that a smaller amount of training data is needed for the models to converge successfully. Therefore, by capitalizing on the conditional independence assumption, the overall training time can be significantly decreased. Additionally, smaller networks result in faster inference runtimes. Hereinafter, we refer to as the qubit corresponding to $C$ as the conditional qubit.

% \begin{figure*}[t]
%     \centering
%         \subfloat[]{
%             \includegraphics[width=0.4\linewidth]{Figure2a.pdf}}
%             \hspace{5mm}
%         \subfloat[]{
%             \includegraphics[width=0.35\linewidth]{Figure2b.pdf}}
%     \caption{ The basic idea of this paper for scalable QMEM. (a) the partitioning of qubits according to the conditional independence assumption. This process allows us to factorize a joint probability distribution in to conditional probability distribution. (b) depicts the process of transfer learning. The error map of the source subsystem is pre-trained using a neural network consisting of fully connected layers. The model depited below is initialized with the pre-trained weights obtained from above. We trained only the last layer to train the target subsystem.}        
% \end{figure*}
\begin{figure}[t]
    \centering
    \includegraphics[width=0.95\linewidth]{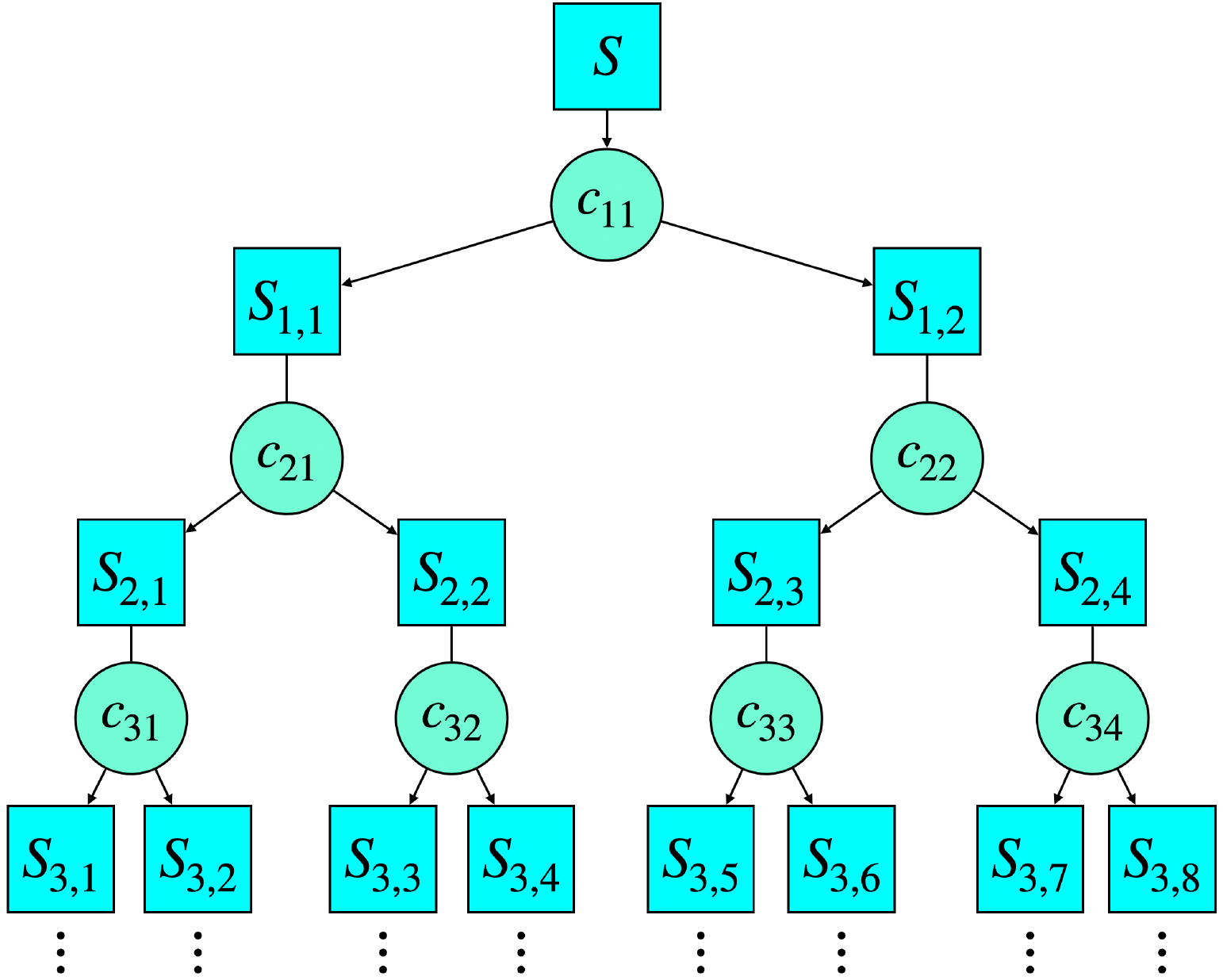}
    \caption{\label{fig:fig2}The partitioning of qubits according to the conditional independence assumption. $S_{i,j}$ represents the $j^{\mathrm{th}}$ subsystem at partition level $i$. $c_{ij}$ represents the conditional qubit that connects the subsystems $S_{i,2j-1}$ and $S_{i,2j}$. This partitioning scheme assumes that the subsystems are independent given the state of $c_{ij}$. This process allows us to factorize a joint probability distribution into conditional probability distributions.}        
\end{figure}
The partitioning of the given quantum system based on the principle of conditional independence is illustrated in Fig.~\ref{fig:fig2}. In this figure, $S_{i,j}$ represents the $j^{\mathrm{th}}$ subsystem at partition level $i$, while $c_{ij}$ denotes the conditional qubit that connects the subsystems $S_{i,2j-1}$ and $S_{i,2j}$. This partitioning scheme assumes that the subsystems are independent given the state of $c_{ij}$. Let us denote a set of conditional qubits that connect a leaf $S_{i,j}$ and the root by $C_{S_{i,j}}$. For example, $C_{S_{3,3}}=\lbrace c_{11}, c_{21}, c_{32} \rbrace$ and $C_{S_{3,6}}=\lbrace c_{11}, c_{22}, c_{33} \rbrace$. The partitioning process continues until the leaf nodes are reached, where each leaf node consists of a small number of qubits (e.g., less than 10). For instance, if the partitioning in Fig.~\ref{fig:fig2} terminates at level 3, we would have eight leaf nodes. Each leaf node is associated with a conditional probability distribution $p(S_{3,i}|C_{S_{3,i}})$, where $i\in\lbrace 1,\ldots,8\rbrace$. The full joint probability distribution under CI is then computed by $\prod_{i=1}^{8}p(S_{3,i}|C_{S_{3,i}}) \prod_{i=1}^{3}\prod_{j=1}^{2^{i-1}}p(c_{ij})$. In this example, each conditional probability distribution requires the training of eight separate neural networks, taking into account all computational basis states of the conditional qubits in $C_{S_{3,i}}$. However, the number of conditional qubits grows with the depth of the tree, which grows logarithmically with the number of total qubits. Moreover, it is evident that the number of leaf nodes cannot exceed the number of total qubits. Therefore, the number of neural networks to be trained independently grows linearly with the number of total qubits. The size of each neural network is constant because the leaf nodes are designed to contain only a small constant number of qubits. This constitutes an efficient QMEM method, which we refer to as CI-QMEM, that is exponentially faster than previous methods that aim to correct for the full joint probability distribution model without conditional independence, for which the size of the neural network or the size of the linear response matrix grows exponentially with the number of qubits.

The general formula for computing the joint probability distribution of a total system $S$ whose partitioning under CI terminates at level $L$ is
\begin{equation}
\label{eq:jointp}
    p(S)=\prod_{i=1}^{2^L}p(S_{L,i}|C_{S_{L,i}}) \prod_{i=1}^{L}\prod_{j=1}^{2^{i-1}}p(c_{ij}).
\end{equation}

%\begin{figure}[t]
%    \centering
%        \includegraphics[width=0.8\linewidth]{Figure2.pdf}
%        \caption{ Partitioning of qubits according to the conditional independence assumption.}
%\end{figure}

% Let $N$ be the number of qubits, $N_{cq}$ be the number of conditional qubits. We assume that one conditional qubit is divided into two partitions. So, the number of partitions is $N_{cq} + 1$ and each partition has $n_{1}, n_{2}, ..., n_{cq + 1}$ qubits. Then, the input node size is as follows:
% \begin{align*}
%     \left( \sum_{i = 1}^{cq + 1} 2^{n_{i}} \right) \times 2^{N_{cq}} + N_{cq} 2^{1}
% \end{align*}
% where $\sum_{i = 1}^{cq + 1} n_{i} = N - N_{cq}$. \\
% In general, we can partition the total system into two (with one conditional qubit), then again partition each subsystems into two (by choosing one conditional qubit in each subsystem), and iterate this process until there are only 3 or 4 qubits left in the subsystem without any independence. In this way, the size of the neural network will be significantly smaller.
%Compared to the full input node size, 
%\begin{align*}
%    2^{N}   &\ge \left( \sum_{i = 1}^{cq + 1} 2^{n_{i}} \right) \times 2^{N_{cq}} + 2^{N_{cq}} \\
%            &= \left( \sum_{i = 1}^{cq + 1} 2^{n_{i}} + 1 \right) \times 2^{N_{cq}}
%\end{align*}
%\begin{align*}
%    2^{N - N_{cq}} = 2^{\sum_{i = 1}^{cq + 1} n_{i}} \ge \sum_{i = 1}^{cq + 1} 2^{n_{i}}  + 1
%\end{align*}
%\begin{align*}
%    \sum_{i = 1}^{cq + 1} n_{i} \log{2} \ge \log\left( {\sum_{i = 1}^{cq + 1} 2^{n_{i}} } + 1 \right) 
%\end{align*} 

\subsection{Transfer Learning}
\label{sec:sec3_sub2}
% Transfer Learning
Transfer learning (TL) can further reduce the training run-time by leveraging pre-trained neural networks. Instead of training a new neural network from scratch on a new dataset, transfer learning involves using parameters of a pre-trained network on a reference dataset that shares some similarities with the new dataset~\cite{taylor2009transfer, tan2018survey, raffel2020exploring}. Typically, the lower (earlier) hidden layers of a neural network, which capture low-level features, are kept frozen, while only the upper (later) layers are fine-tuned or trained. This strategy eliminates the need to relearn the common low-level features, enabling faster convergence and reducing the overall training time. 

\begin{figure}[t]
    \centering
    \includegraphics[width=0.95\linewidth]{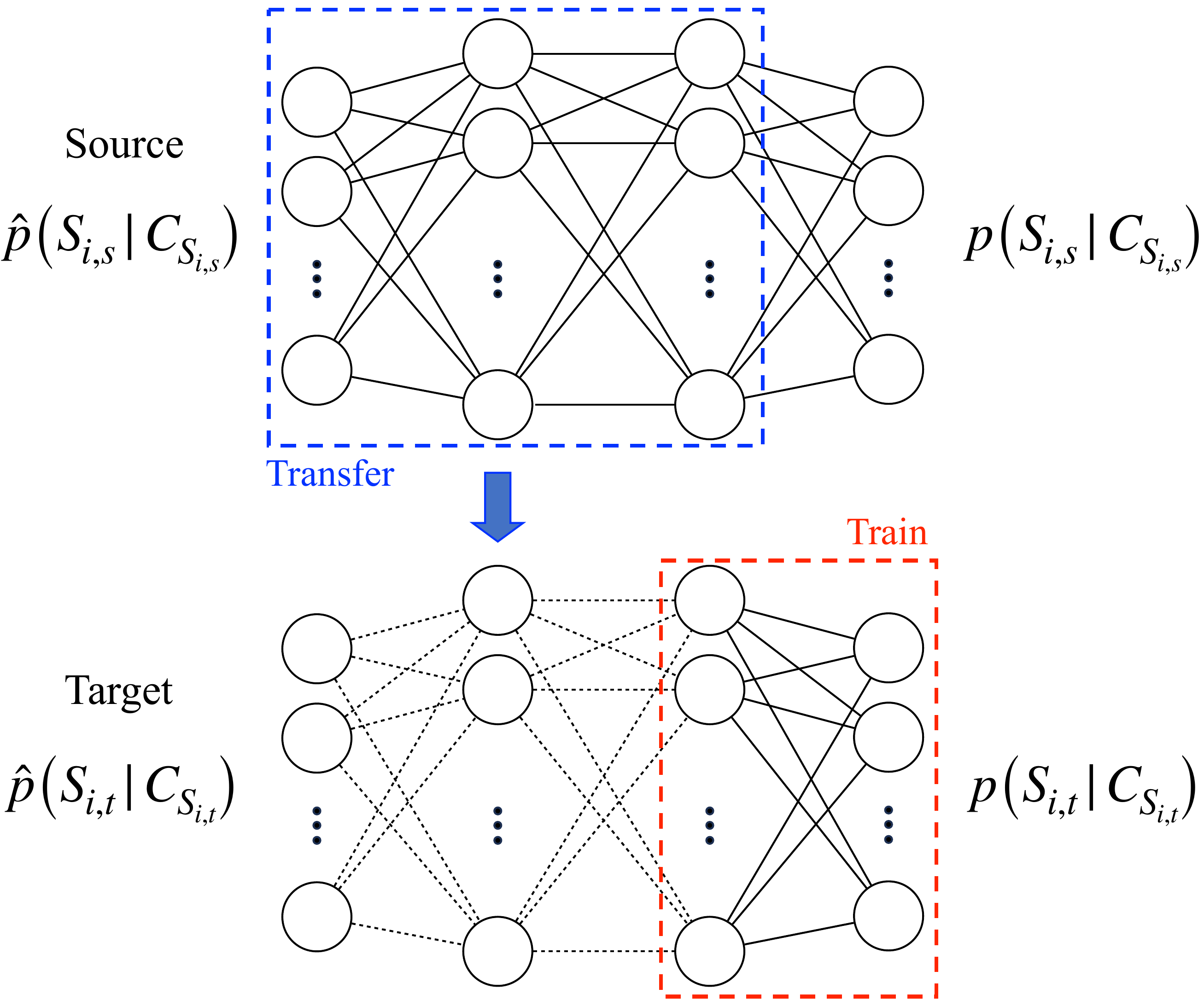}
    \caption{\label{fig:fig3}Transfer learning process for training neural networks in CI-QMEM. Initially, a neural network is trained to correct measurement errors of the source subsystem, denoted by $S_{i,s}$. The training utilizes the noisy probability distribution, $\hat{p}(S_{i, s} | C_{S_{i, s}})$ and  $p(S_{i, s} | C_{S_{i, s}})$, as the input and the output, respectively. The parameters within the hidden layers, enclosed by a blue dashed box in the source model, are transferred to a new neural network designed to address measurement errors in a target subsystem, $S_{i,t}$. This transfer is achieved by initializing the dashed edges in the target neural network with their corresponding parameters from the source neural network. Subsequently, only the last hidden layer of the target model, represented by a red dashed box, undergoes training.}        
\end{figure}
To understand how TL works in QMEM, let us consider a source subsystem denoted as $S_{i, s}$ with its associated conditional qubits represented by $C_{S_{i, s}}$. CI-QMEM trains a neural network for each computational basis state of the conditional qubits $C_{S_{i, s}}$ by utilizing the noisy and ideal conditional probability distributions, $\hat{p}(S_{i, s} | C_{S_{i, s}})$ and $p(S_{i, s} | C_{S_{i, s}})$, as input and output, respectively. Now, consider a target subsystem $S_{i, t}$ with the same number of qubits as $S_{i, s}$, also having its associated conditional qubits denoted by $C_{S_{i, t}}$. One might initially attempt repeat the entire CI-QMEM procedure described above to train a neural network for this new system. However, if the noise characteristics experienced by these subsystems exhibit similarities that can be captured by the neural network trained for the source system, it is unnecessary to train a separate neural network for the target system from the beginning. Instead, it is possible to transfer selected parameters learned from the source system to the target system, thereby reducing the number of parameters that need to be trained. Therefore, the application of TL is a justifiable approach, particularly when assuming the presence of systematic sources of noise that are common across different qubits within the quantum device. This concept is illustrated in Fig.~\ref{fig:fig3}. Importantly, a source model can be utilized for multiple target subsystems, as long as they share similar features with the source. Transfer learning typically reduces the number of parameters subject to training in the target model by a constant amount. Consequently, the reduction in trainable parameters also leads to a decrease in the amount of training data required for the model to converge. Henceforth, we refer to the QMEM technique that combines both CI and TL as CITL-QMEM.

\section{Implementation}
\label{sec:section4}
\subsection{Data collection}
\label{sec:sec4_sub1}

Deep learning-based QMEM methods require sample data for training neural networks. When constructing a family of quantum circuits to generate the training dataset, an essential requirement is the efficient computation of the associated probability distribution on a classical computer. This is necessary because training involves both noisy and ideal measurement results. Moreover, it is crucial that the error introduced by the gates used to prepare quantum states for the training data is negligible compared to the measurement error.  In modern quantum devices, single-qubit gate errors are typically insignificant compared to measurement errors. Therefore, using quantum circuits composed solely of single-qubit gates satisfies these conditions. 

Since the objective of QMEM is to mitigate errors in projective measurements in the computational basis, defined by the eigenstates of $\sigma_z$, the relative phase between computational basis states is irrelevant. Consequently, quantum circuits that generate training data employ only single-qubit $R_{y}(\theta)$ gates. Specifically, the unitary operation preparing the state is $\prod_{i=1}^{n}R^{(i)}{y}(\theta_i)$, where the superscript $(i)$ denotes that the single-qubit rotation is applied to the $i^{\mathrm{th}}$ qubit. The rotation angles are sampled randomly in a way that resulting quantum states are distributed uniformly on the boundary of the $xz$-plane of the Bloch sphere.
This is achieved by randomly generating values for $z_{i} \in [-1, 1]$ and computing $\theta_{i} = \arccos(z_{i})$. The noisy probability distribution $\hat{\boldsymbol{p}}$ generated by the quantum circuit serves as the input for the neural network, while the ideal probability distribution $\boldsymbol{p}$ represents the desired output. The computation of $\boldsymbol{p}$ can be expressed as follows:
\begin{align*}
    p(b) = \left| \prod_{i}^n \cos^{1-b_{i}}(\theta_{i}/2) \sin^{b_{i}}(\theta_{i}/2) \right|^2,
\end{align*}
where $b_{i}$ is the $i^{\mathrm{th}}$ bit of the binary string $b$.

In the LI-QMEM method, the training data comprises all computational basis states of the target qubit system, requiring $2^n$ circuit executions for an $n$-qubit system.  The data acquisition and error mitigation process can be facilitated using the Qiskit Ignis package~\cite{anis2021qiskit}. This package constructs a $2^n \times 2^n$ calibration matrix for the $n$-qubit system based on $2^n$ pairs of noisy and ideal results.
%%%% Edited up to here - DKP
\subsection{Model construction}
\label{sec:sec4_sub2}
%To train a neural network (NN), we need a training data set, define the NN architecture (hidden layers, nodes, activation functions), select a loss function to measure performance on training data, and an optimization algorithm to update weights and biases during training.  

% The training data was generated using a quantum circuit by applying single-qubit $R_{y}(\theta)$ gate to each qubit and measuring in the computational basis, assuming that the single-qubit gate error is negligible compared to that of the measurement error. The rotation angles are sampled so that the resulting quantum states spread evenly on the boundary of the $xz$-plane of the Bloch sphere.
% %The impact of single-qubit gate is considered negligible compared to that of two-qubit gates and measurements. To ensure uniformity of the data for projective measurements in the computational basis, random values $z_{i} \in [-1, 1]$ were generated, and calculated $\theta_{i}$ using the equation  $\theta_{i} = \arccos(z_{i})$. 

% %The observed probability vector $\hat{\boldsymbol{p}}$ from the quantum circuit was used as the input for the NN, with the ideal probability vector $\boldsymbol{p}$ as the output.
The neural network architecture comprises multiple layers, including the input, hidden, and output layers. For a joint probability distribution involving $n$ qubits, there exist $2^{n}$ possible computational basis states. Hence, both the input and output layers consist of $2^{n}$ nodes. To achieve an optimal balance between convergence speed and accuracy, thorough experimentation was conducted to select appropriate hyperparameters. Specifically, we configured the network with 4 hidden layers, and each hidden layer contained $5 \times 2^{n}$ nodes. All hidden layers are fully connected layers, and each hidden node employs the Scaled Exponential Linear Unit (SELU) as the activation function~\cite{NIPS2017_5d44ee6f,PhysRevApplied.17.014024}. The output layer employs the softmax activation function, which normalizes the outputs into a probability distribution, ensuring that the sum of the output values is equal to one. The weights and biases of the neural network are optimised using the categorical cross-entropy loss function. The parameters are updated by the Adam optimizer~\cite{kingma2014adam}. As part of the hyperparameter tuning, we set the learning rate, batch size, and number of epochs to 0.0001, 16, and 300, respectively.

\subsection{Experimental Results}
\label{sec:sec4_sub4}

\begin{figure*}[ht]
    \centering
    \subfloat[ibmq\_jakarta]{
        \includegraphics[width=0.405\textwidth]{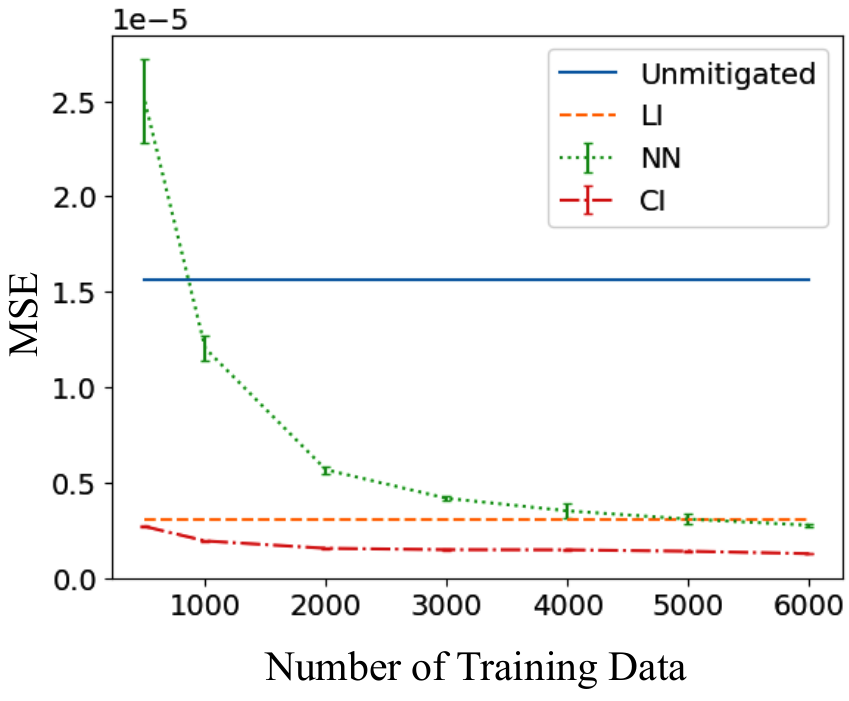}
        \label{fig:first}
    }
    \hspace{5mm}
    \subfloat[ibm\_lagos]{
        \includegraphics[width=0.41\textwidth]{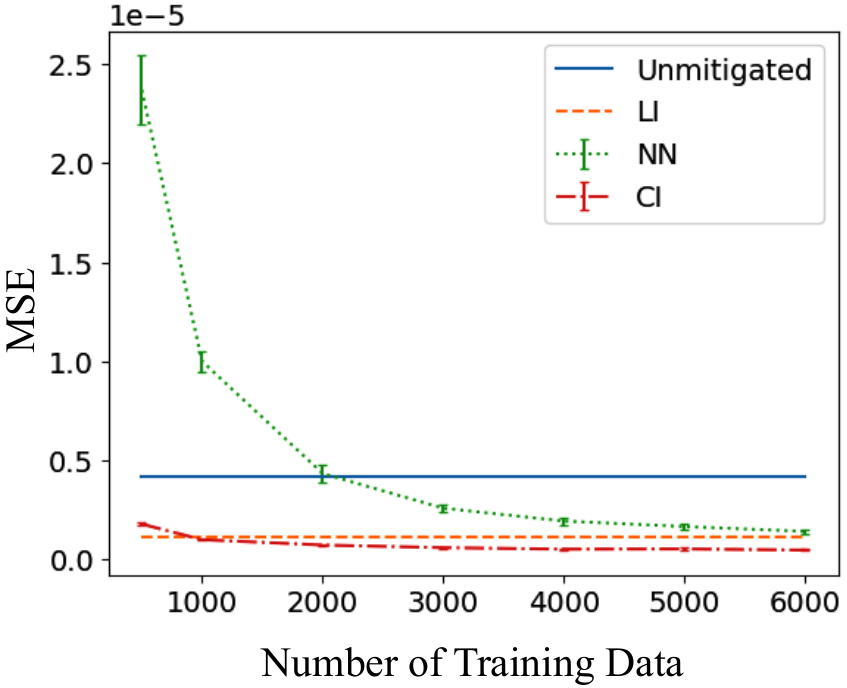}
        \label{fig:second}
    }
    \caption{\label{fig:fig4} Experimental results showing the mean squared error (MSE) between the ideal and error-mitigated probability distributions as a function of the number of training data. In the figure legend, we label the results from LI-QMEM, NN-QMEM, and CI-QMEM as LI, NN, and CI, respectively. The experiments were conducted on two 7-qubit devices: (a) ibmq\_jakarta and (b) ibm\_lagos. To evaluate the performance, 1500 test data points were used after training the model on varying numbers of training data. Each experiment was repeated five times, with the data set randomly split into training and test sets for each run. For NN and CI, the plotted values represent the mean MSE, with the error bars indicating one standard deviation. For the unmitigated and LI results, the smallest MSE obtained across the five experiments is shown.}
\end{figure*}

%To verify the effectiveness of our scalable QMEM in reducing errors efficiently, we compare it with LI-QMEM and NN-QMEM. LI-QMEM used the calibration matrix obtained using the CompleteMeasFitter from the qiskit Ignis package, and NN-QMEM used a neural network architecture as described in Section~\ref{sec:sec4_sub2}. First, we conducted experiments using IBM's 7-qubit quantum devices, ibmq\_jakarta and ibm\_lagos. To train the deep learning-based QMEM method, we generated 7,500 data for each device and split them into 6,000 for training and 1,500 for testing. To perform LI-QMEM, we generated $2^7$ calibration circuits to obtain the calibration matrix. The joint probability distribution of the 7-qubit quantum system was decomposed at level 1 according to Equation (1), as $p(q_{0},q_{1},q_{2}|q_{3}) p(q_{4},q_{5},q_{6}|q_{3}) p(q_{3})$, and for each conditional probability distribution, we need two separate neural network, considering all the computational basis states of the conditional qubit $q_{3}$. In total, we trained $2 \times 2 + 1$ neural networks. 

To validate the effectiveness of the proposed QMEM methods and to compare their performances against existing ones, we conducted experiments on 7-qubit and 13-qubit quantum systems. We assessed the performance of different QMEM methods using three metrics: Mean Squared Error (MSE), Kullback-Leibler Divergence (KLD), and Infidelity (IF). These metrics quantify the dissimilarity between the ideal and mitigated probability distributions and are computed as follows:
\begin{align*}
    D_{\mathrm{MSE}} = {1 \over 2^{n}} \sum_{i=1}^{2^{n}} \left| \tilde{p}_{i} - p_{i} \right| ^2,
\end{align*}
\begin{align*}
    D_{\mathrm{KLD}} = \sum_{i=1}^{2^{n}} p_{i} \log{{p_{i}} \over {\tilde{p}_{i}}} ,
\end{align*}
\begin{align*}
    D_{\mathrm{IF}} = 1 - \left( \sum_{i=1}^{2^{n}} \sqrt{p_{i} \tilde{p}_{i}} \right) ^2.
\end{align*}
Here, $p_i$ and $\tilde{p}_{i}$ represent the $i^{\mathrm{th}}$ elements of the vectors representing the ideal and the mitigated probability distributions, respectively. A lower value for these measures indicates better performance, as it signifies a closer match between the ideal and mitigated distributions. Additionally, to quantify the rate of error reduction, we use the rate of improvement for each loss function $D_{i}$, where the subscript $i$ corresponds to MSE, KLD, or IF. This rate of improvement, denoted as $R_{i}$, is calculated as follows~\cite{kim2022quantum}:
\begin{equation*}
    R_{i} = \frac{D_i^{\text{Unmitigated}} - D_{i}^{\text{QMEM}}}{D_i^{\text{Unmitigated}}} \times 100(\%).
\end{equation*}
A higher value of $R_{i}$ indicates better performance in reducing errors using QMEM.

The 7-qubit experiments were conducted using two IBM quantum devices, namely ibmq\_jakarta and ibm\_lagos. To obtain the calibration matrix for LI-QMEM, we generated a set of $2^7$ calibration circuits. For training the deep learning-based QMEM method, we generated 7500 data points for each device, which were then split into 6000 samples for training and 1500 samples for testing. To estimate the probability distribution, each quantum circuit was repeated $3.2\times 10^4$ times.

\begin{figure*}[ht]
    \centering
    \subfloat[MSE]{
        \includegraphics[width=0.3\textwidth]{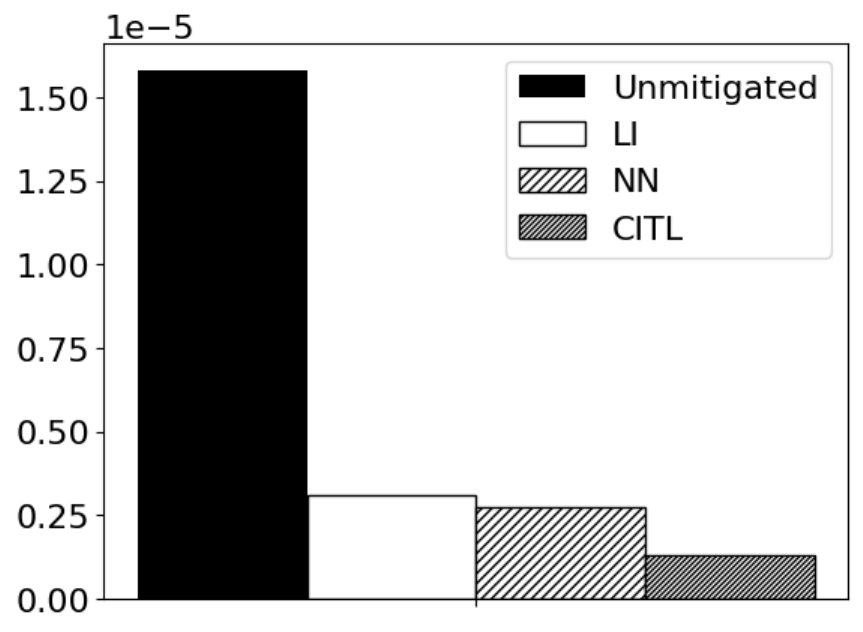}
        \label{fig:first}
    }
    \subfloat[KLD]{
        \includegraphics[width=0.3\textwidth]{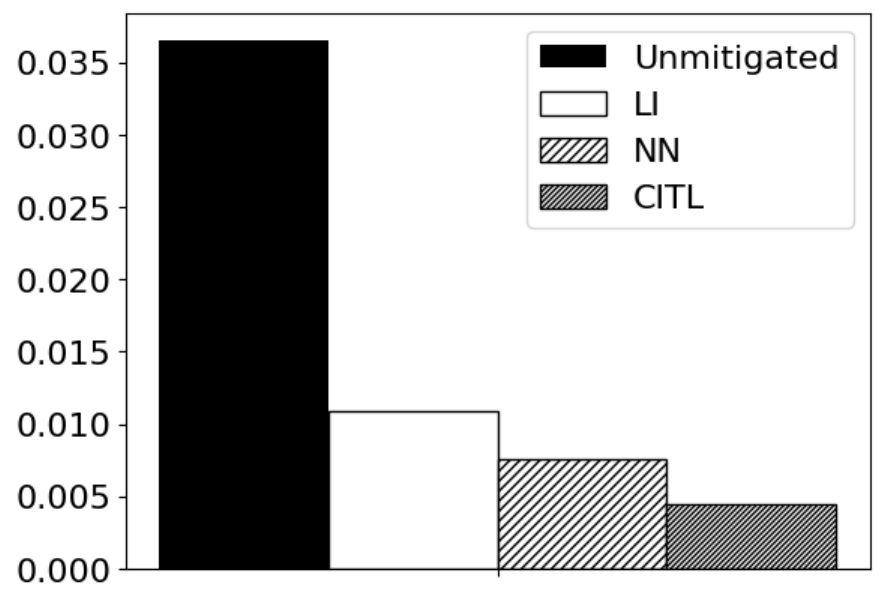}
        \label{fig:second}
    }
    \subfloat[Infidelity]{
        \includegraphics[width=0.3\textwidth]{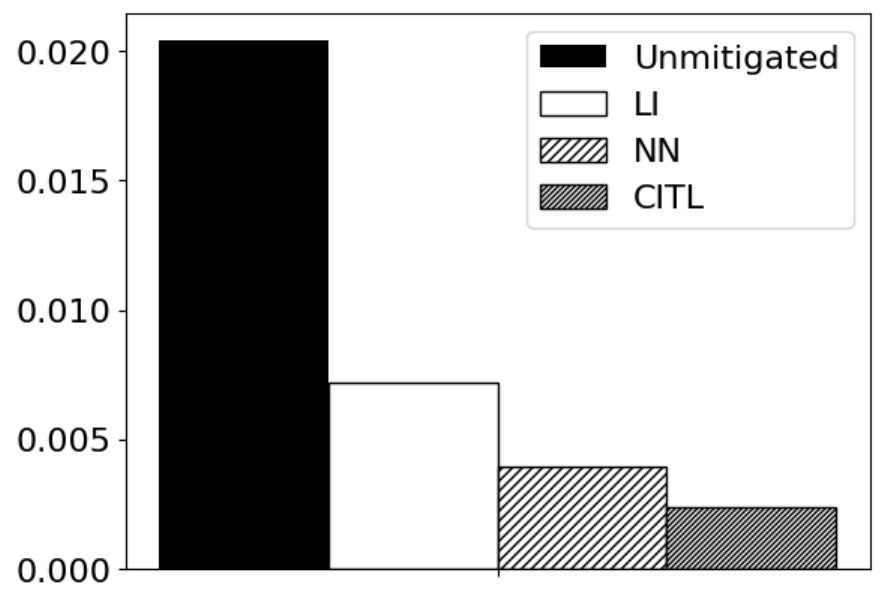}
        \label{fig:second}
    }
    \caption{\label{fig:fig5} To compare the 7-qubit QMEM results obtained using different methods on the ibmq\_jakarta, we measured (a) MSE, (b) KLD, and (c) IF. The filled bars represent the unmitigated results, the unfilled bars represent the results obtained using the LI-QMEM method, the lighter hatched bars (third from the left) represent the results obtained using the NN-QMEM method, and the darker hatched bars (fourth from the left) represent the results obtained using the CITL-QMEM method.
}
\end{figure*}
\begin{figure*}[ht]
    \centering
    \subfloat[MSE]{
        \includegraphics[width=0.28\textwidth]{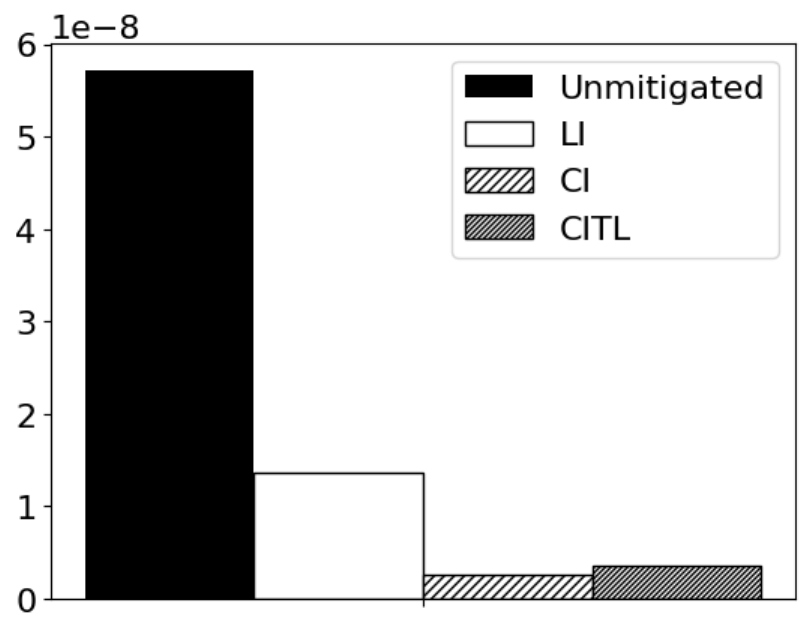}
        \label{fig:first}
    }
    \subfloat[KLD]{
        \includegraphics[width=0.3\textwidth]{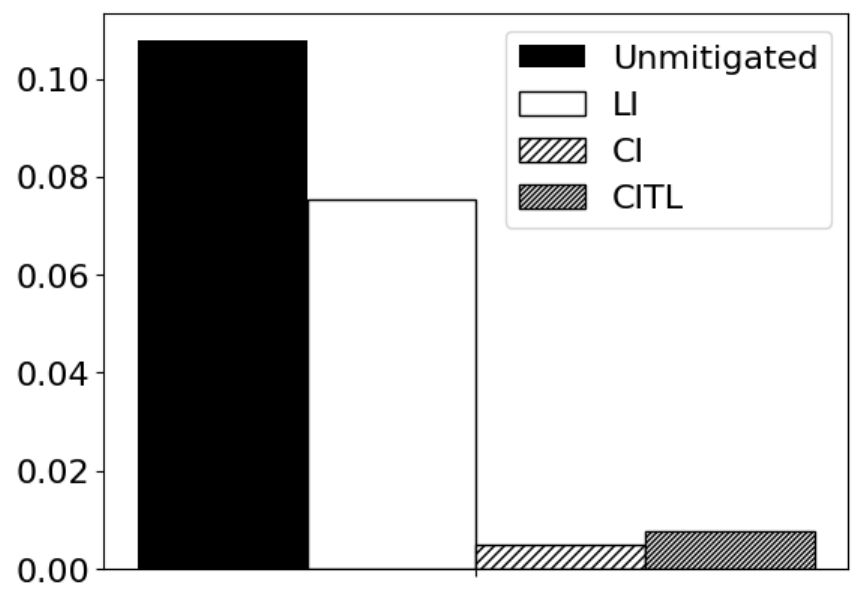}
        \label{fig:second}
    }
    \subfloat[Infidelity]{
        \includegraphics[width=0.3\textwidth]{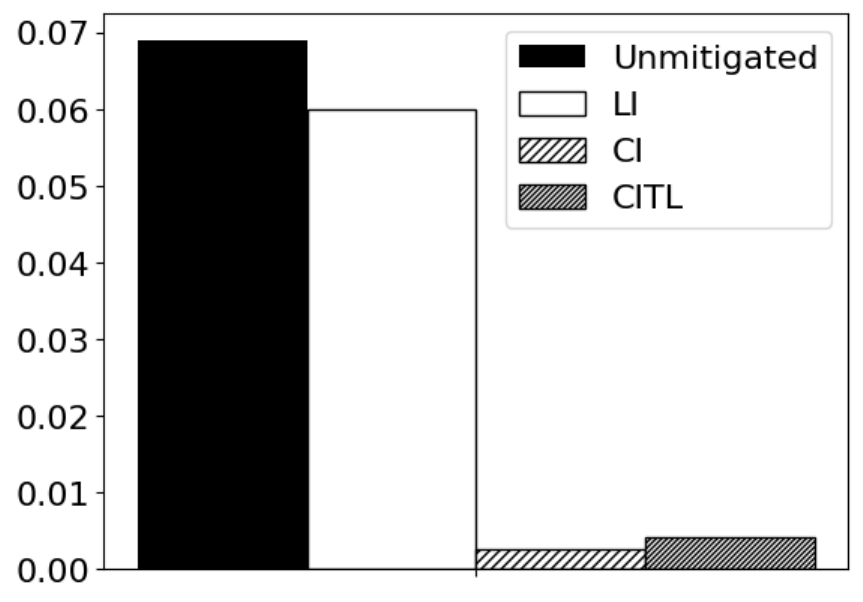}
        \label{fig:second}
    }
    \caption{\label{fig:fig6} To compare the 13-qubit QMEM results obtained using different methods on the ibmq\_kolkata, we measured (a) MSE, (b) KLD, and (c) IF. The filled bars represent the unmitigated results, the unfilled bars represent the results obtained using the LI-QMEM method, the lighter hatched bars (third from the left) represent the results obtained using the CI-QMEM method, and the darker hatched bars (fourth from the left) represent the results obtained using the CITL-QMEM method.
}
\end{figure*}
\setlength{\tabcolsep}{6pt} % Default value: 6pt
\renewcommand{\arraystretch}{1} % Default value: 1
\begin{table}[]
    \begin{tabular}{lccccc}
    \toprule
                    &                    & LI    & NN    & CI    & CITL         \\ \midrule
                    & $R_{\mathrm{MSE}}$ & 80.47 & 82.56 & 91.76 & 91.91        \\
    ibmq\_jakarta   & $R_{\mathrm{KLD}}$ & 70.08 & 79.23 & 87.80 & 87.21        \\
                    & $R_{\mathrm{IF}}$  & 64.68 & 80.64 & 88.24 & 87.54        \\ \midrule
                    & $R_{\mathrm{MSE}}$ & 72.02 & 63.27 & 87.96 & 89.32        \\
    ibm\_lagos      & $R_{\mathrm{KLD}}$ & 53.67 & 64.57 & 83.52 & 85.13        \\
                    & $R_{\mathrm{IF}}$  & 43.29 & 66.67 & 84.20 & 85.76        \\ \bottomrule
    \end{tabular}
    \caption{\label{table:table1} Performance evaluation of each QMEM method on the 7-qubit quantum devices, represented by the rate of improvement ($R_i$). The results for LI-QMEM, NN-QMEM, CI-QMEM, and CITL-QMEM are displayed in the first, second, third and fourth columns, respectively.}
\end{table}

The results of the 7-qubit QMEM experiments for evaluating the MSE between the ideal and error-mitigated probability distributions as a function of the number of training data are shonw in Fig.~\ref{fig:fig4}. The results indicate that CI-QMEM achieves substantial error mitigation using less data compared to NN-QMEM. This observation is consistent with the well-known fact that the amount of data required for training deep neural networks depends on the model complexity~\cite{hoge2018primer, alzubaidi2021review, hu2021model, sarker2021deep}. As CI-QMEM significantly reduces the size of neural networks and the number of model parameters, it improves the error mitigation performance, even with fewer data. Specifically, the conventional NN-QMEM trains a neural neural network with about $1.8\times 10^6$ parameters, while CI-QMEM uses only about $2.9\times 10^4$ parameters. The LI-QMEM, despite using only 128 data, cannot effectively mitigate non-linear errors as the deep learning-based approaches can. Consequently, CI-QMEM achieves smaller measurement errors, underscoring its superior performance in error mitigation compared to other methods. Based on the observation that the improvement from LI-QMEM to CI-QMEM is more pronounced in ibmq\_jakarta compared to ibm\_lagos, we speculate that the non-linear error is stronger in the former than in the latter. Moreover, it is important to note that the number of data required in LI-QMEM grows exponentially with the number of qubits.

CITL-QMEM, which incorporates TL in addition to CI, achieves a further reduction in the number of parameters to about $1.5\times 10^4$. This is accomplished by selecting the neural network trained for learning the conditional probability distribution, $p(q_{0},q_{1},q_{2}|q_{3})$, as the source model. Subsequently, the target model learns the new conditional probability distribution, $p(q_{4},q_{5},q_{6}|q_{3})$. In this TL approach, only the last layer of the target neural network is fine-tuned (trained), while the rest of the hidden layers retain the parameters from the source model.

The overall results for the 7-qubit QMEM are reported in Table ~\ref{table:table1}, and Fig.~\ref{fig:fig5} illustrates the effectiveness of transfer learning in reducing errors. Since the amount of error mitigated by CI-QMEM and CITL-QMEM is comparable as shown in the table, the figure only presents the results from the latter.

\setlength{\tabcolsep}{6pt} % Default value: 6pt
\renewcommand{\arraystretch}{1} % Default value: 1
\begin{table}[]
    \begin{tabular}{lccccc}
    \toprule
                    &                    & LI    & CI    & CITL         \\ \midrule
                    & $R_{\mathrm{MSE}}$ & 51.35 & 93.65 & 94.40        \\
    ibmq\_mumbai    & $R_{\mathrm{KLD}}$ & -20.82& 92.09 & 92.25        \\
                    & $R_{\mathrm{IF}}$  & -31.75& 94.21 & 94.34        \\ \midrule
                    & $R_{\mathrm{MSE}}$ & 76.01 & 95.38 & 93.68        \\
    ibm\_kolkata    & $R_{\mathrm{KLD}}$ & 30.11 & 95.53 & 92.78        \\
                    & $R_{\mathrm{IF}}$  & 13.14 & 96.43 & 93.91        \\ \bottomrule
    \end{tabular}
    \caption{\label{table:table2} Performance evaluation of each QMEM method on the 13-qubit quantum devices, represented by the rate of improvement ($R_i$). The results for LI-QMEM, CI-QMEM, and CITL-QMEM are displayed in the first, second, and third columns, respectively.}
\end{table}

The 13-qubit QMEM experiments were conducted using two IBM quantum devices, namely ibmq\_mumbai and ibmq\_kolkata. These devices consist of 27 qubits, and for our experiment, we selected 13 of them as shown in Fig.~\ref{fig:fig1_2}. We generated 6000 data for each device and split them into 5950 for training and 50 for testing. To estimate the probability distribution, we repeated the quantum circuit $1.0\times 10^5$ times, which is the maximum number available as the cloud service allows. 

As shown in Fig.~\ref{fig:fig1_2}. The joint probability distribution for the 13-qubit system decomposes as $\prod_{i=1}^{2^{2}}p(S_{2,i}|C_{S_{2,i}}) \prod_{i=1}^{2}\prod_{j=1}^{2^{i-1}}p(c_{ij})$, where $S_{2,1}=\lbrace q_{4},q_{6},q_{7}\rbrace$, $S_{2,2}=\lbrace q_{12},q_{15}\rbrace$, $S_{2,3}=\lbrace q_{5},q_{8},q_{9}\rbrace$, $S_{2,4}=\lbrace q_{14},q_{16}\rbrace$, $c_{11} = q_{13}$, $c_{21} = q_{10}$, and $c_{22} = q_{11}$. Each conditional probability distribution requires training four separate neural networks, taking into account all computational basis states of the conditional qubits. Consequently, the total number of neural networks to be trained is $4 \times 4 + 3 = 19$. For transfer learning, we selected $S_{2,1}$ and $S_{2,2}$ as source subsystems. The results are presented in Fig.~\ref{fig:fig6} and Table~\ref{table:table2}. For the 13-qubit system, we did not perform conventional NN-QMEM experiments as it requires training a neural network with about $7.3$ billion parameters, leading to unreasonably long data collection and training times. As evident from both the figures and the table, our methods significantly reduce errors in all performance measures.  Interestingly, on the ibmq\_mumbai device, CITL-QMEM performs slightly better than CI-QMEM. This can be attributed to the use of a pre-trained model, enabling the creation of a simpler neural network with enhanced generalization capability~\cite{yosinski2014transferable, alzubaidi2020towards}. It is important to note that while LI-QMEM requires 8192 data points and NN-QMEM requires 7.3 billion parameters, our method is trained with only 5950 data points and $7.4\times 10^4$ parameters for CI-QMEM, and $3.9\times 10^4$ parameters for CITL-QMEM.

\section{Conclusions and Discussion}
\label{sec:section5}
We introduced a scalable quantum measurement error mitigation method that overcomes the limitations of existing approaches. By utilizing conditional independence and transfer learning techniques, we achieve exponential reductions in the size of neural networks while maintaining excellent error-mitigation capabilities. Our method not only reduces the size of neural networks and the number of parameters to optimize but also significantly decreases the amount of data required for effective training. Experimental results on four IBM quantum devices, featuring 7 and 13 qubits, provide strong evidence of the efficiency and effectiveness of our method in mitigating measurement errors. Notably, CI-QMEM demonstrates a substantial reduction of neural network parameters by approximately 60 times for 7-qubit systems and $10^5$ times for 13-qubit systems. Transfer learning allows for the additional reduction of parameters by approximately a factor of two in both cases, leading to even more efficient training. As a result, CI-QMEM and CITL-QMEM consistently outperformed the full NN-QMEM under similar training conditions in all experiments. Moreover, these deep learing-based methods outperformed the LI-QMEM due to their ability to correct non-linear errors. In particular, for the 13-qubit experiments, the enhancements were achieved while using a smaller amount of data.

In the following, we discuss interesting future research directions and open problems. While the primary focus of this study is to mitigate measurement errors, the techniques developed here can be combined with existing methods for mitigating gate errors~\cite{temme2017error,endo2018practical,PhysRevA.98.013414,9226505,kurita2022synergetic}, thereby enhancing overall quantum computing performance. Integrating gate error mitigation approaches with measurement error mitigation techniques holds great potential for significant improvements in the accuracy and reliability of quantum computations. Moreover, it would be intriguing to extend the concepts of CI and TL explored in this work to improve the efficiency and scalability of the existing machine learning-based gate error mitigation technique~\cite{9226505}. The QMEM methods discussed in this work operate on the probability distribution obtained as the final outcome of a quantum computation at the software level. Consequently, these approaches are particularly suitable for cloud-based quantum computing environments. An interesting future endeavor is to optimize the performance by combining these approaches with hardware-level techniques that specifically addresses improving qubit-state-assignment fidelity by working directly with the readout signals~\cite{PhysRevApplied.17.014024}. Furthermore, the success of CI-QMEM implies that the underlying assumption of negligible measurement cross-talk among distant qubits holds. Exploring the reverse scenario and developing techniques to characterize measurement crosstalk by leveraging the concept of conditional independence presents an interesting avenue for future research. An important open problem in the general NN-QMEM, including our approach, pertains to the potential impact of shot noise in large systems. For instance, when working with training data that is close to a uniformly-distributed state, the number of shots required to obtain its noisy distribution grows exponentially with the number of qubits. Future research can explore the effectiveness of CI and TL based QMEM on a biased training set, where the smallest probability is in $O(1/\mathrm{poly(n)})$, or develop tailored methods for such conditions.

% \section*{Data Availability}
% The data that support the findings of this study are available upon request.

\section*{Acknowledgments}

This research was supported by the Yonsei University Research Fund of 2023 (2023-22-0072), the National Research Foundation of Korea (Grant No. 2022M3E4A1074591), and the KIST Institutional Program (2E32241-23-010).

% \bibliographystyle{unsrt}
% \bibliography{ref}

\end{document}